\documentstyle[12pt]{article}
\input epsf.sty
 1

\catcode`\@=11 % This allows us to modify PLAIN macros.
\makeatletter
\def\@seccntformat#1{\csname the#1\endcsname.\hskip 1em}
\makeatother
\begin{document}

%%%%%%%%%%%%%%%%%%%%%%%%%%%%%%%%%%%%%%%%%%%%%%%%%%%%%%%
%------------------------------------------------------
% Title page
%------------------------------------------------------
\thispagestyle{empty}
\begin{flushright}

{\footnotesize\renewcommand{\baselinestretch}{.75}
           SLAC-PUB-8046\\
           December 1998\\
}
\end{flushright}

\vspace {0.5cm}

\begin{center}
{\large \bf  PHYSICS RESULTS FROM SLD USING THE CRID$^*$}

\vspace {1.0cm}

 {\bf David Muller}

\vspace {0.2cm}

 {\bf Representing The SLD Collaboration$^{**}$}

\vspace {0.2cm}

Stanford Linear Accelerator Center \\
Stanford University, Stanford, CA~94309 \\

\vspace{1.3cm}
{\bf Abstract}
\end{center}

\renewcommand{\baselinestretch}{1.2}

We review recent $Z^0$ physics results from SLD that use the Cherenkov Ring
Imaging Detector for charged particle identification.
The performance of the detector and likelihood method are described briefly.
Several hadronization measurements are presented, including identified hadron
production in events of different primary flavors, leading hadron production,
and new correlation studies sensitive to details of both leading and nonleading
hadron production.
Identified $K^\pm$ have been used in conjunction with precision vertexing to
study charmless and doubly charmed $B$-hadron decays.
This combination has also been used to tag $b$, $\bar{b}$, $c$ and $\bar{c}$
jets, yielding precise measurements of $B^0$-$\bar{B}^0$ mixing and of the
asymmetric couplings $A_b$ and $A_c$.
Identified $K^\pm$ and $\Lambda^0$/$\bar{\Lambda}^0$ have been used to tag $s$
and $\bar{s}$ jets, yielding a measurement of $A_s$.
The clean identified particle samples provided efficiently by the CRID allow the
purities of these tags to be measured from the data, an essential ingredient for
precision physics.

\vspace{1.3cm}
\begin{center}
{\it Presented at the 3 $^{rd}$ International Workshop on Ring Imaging Cherenkov
Detectors,\\
15--20 November 1998, Ein-Gedi, Israel.}
\end{center}

\vfil

\noindent
$^*$Work supported in part by Department of Energy contract DE-AC03-76SF00515.
\eject

\section{Introduction}

The SLD experiment \cite{sld} studies $Z^0$ bosons
produced in $e^+e^-$ annihilatons at the SLAC Linear Collider (SLC).
A carrier of the electroweak interaction, the $Z^0$ boson decays
into a fermion-antifermion ($f\bar{f}$) pair with probability predicted by the
Standard Model (SM) electroweak couplings of the $Z^0$ to fermion $f$.
The parity violation in this decay leads to an asymmetric distribution of the
polar angle between the outgoing $f$ and the incoming $e^-$, which
depends strongly on the $e^+$ and $e^-$ polarizations.
The SLC electron beam is longitudinally polarized to a magnitude of $\sim$73\%
with a sign determined randomly for each beam pulse.

A key aspect of the SLD physics program is the measurement of total and
asymmetric couplings, $R_f$ and $A_f$, for as many of the fundamental fermions $f$ as
possible.
Measuring $A_f$ requires both identifying $Z^0\rightarrow f\bar{f}$ events and
determining the direction of the outgoing $f$ (as opposed to $\bar{f}$),
which is challenging for the quarks, $f=u,d,s,c,b$, as they appear as jets of
particles.
$Z^0\rightarrow b\bar{b}$ and $c\bar{c}$ events can be identified by modern
vertex detectors, using the 3 (1) mm average flight distance of the leading $B$
($D$) hadron in each $b$ ($c$) jet.
Quantities used to distinguish $b$ ($c$) from $\bar{b}$ ($\bar{c}$) jets
include the total charge of vertices or jets, and the charge of identified
leptons or reconstructed $D$ mesons.
The first two methods suffer from low analyzing power, and the other two from
low efficiency.
The charge of identified kaons is foreseen as a powerful method in future $B$
physics experiments, and we have recently pioneered its use for both $b$ and $c$
jets using our Cherenkov Ring Imaging Detector.
The purity of a flavor tag can be measured from the data in $e^+e^-$
annihilations using the anticorrelation between the $f$ and $\bar{f}$ in the 
event.

Light flavor ($u$, $d$, $s$) jets are identifiable using their leading
particles, and early work in this area is promising.
We have used high-momentum strange particles to tag $s$ and $\bar{s}$ jets,
measured the tag purities from the data, and made a measurement of $A_s$.
We are studying ways to saparate $u$, $\bar{u}$, $d$ and $\bar{d}$ jets.
The tagging of light flavors would have a wide range of applications in
high energy physics, from deep inelastic scattering, to jets from
hadron-hadron collisions and studies of the decays of $W$-bosons, top quarks,
Higgs bosons, and any new particle that is discovered.

In addition, the hadronic event samples at the $Z^0$ are of unprecendented size
and purity, providing a unique opportunity to study the structure of hadronic
jets and the decays of $B$ and $D$ hadrons in great detail.
Jet formation is in the realm of non-perturbative QCD and is not understood
quantitatively.
The empirical understanding of jet structure is essential as jets are (will be)
part of the signal for decays of $W^\pm$ bosons, $t$ quarks (any undiscovered
heavy objects), as well as the background for these and other processes.
Isolated jets have been identified with partons in order to make a number of
tests of perturbative QCD.
QCD tests, searches for new physics, and conventional physics studies
would be more sensitive if we could distinguish jets of different origins
(gluons, quarks, antiquarks).
Jet structure in terms of inclusive properties of charged tracks
has been studied extensively, however more theoretical and experimental input is
needed, especially in the area of identified and reconstructed particles.
In particular, the study of leading particles
is needed for the development of light-flavor jet tags.

Decays of $B^0$ and $B^+$ mesons have been studied extensively by
experiments operating at the $\Upsilon$(4S).  The properties and mixture of $B$
hadrons produced at the $Z^0$ differ considerably from those at the
$\Upsilon$(4S), providing a number of opportunities for complementary studies
of $B$ hadron decays \cite{marco}, especially those involving identified
particles for which the $\Upsilon$(4S) experiments have limited momentum
coverage.

To study this wide range of physics, the SLD includes a Cherenkov Ring
Imaging Detector (CRID) designed to identify $\pi^\pm$, $K^\pm$ and
p/$\bar{\rm p}$ over most of the momentum range, and leptons
at low momentum, complemeting the electromagnetic calorimeter and muon
detectors.
The CRID design and performance are summarized in sec. 2.
We then present a number of physics results in the areas of jet structure
(sec. 3) and physics with flavor-tagged jets (sec. 4).
Two unique studies \cite{bdecay} of $B$-hadron decays that benefit from SLD's
excellent vertexing and particle identification are described separately in
these proceedings \cite{marco}.

\section{SLD CRID Performance}

The SLD CRID design and hardware performance are described in \cite{crid}.
Briefly, it is a large barrel detector covering the polar angle range
$|\cos\theta|<0.68$, and comprising two radiator systems; liquid
C$_6$F$_{14}$ and gaseous C$_5$F$_{12}+$N$_2$ cover the lower and higher
momentum regions, respectively.
Cherenkov photons from the liquid (gaseous) radiator are focussed by proximity
(spherical mirrors) onto one of 40 quartz-windowed time projection chambers
(TPCs) containing ethane with $\sim$0.1\% TMAE.
Each single photoelectron drifts to a wire chamber where its conversion point
is measured in three dimensions and used to reconstruct a Cherenkov angle
$\theta_c$ with respect to each extrapolated track.

The average $\theta_c$ resolution for liquid (gas) photons was measured
to be 16 (4.5) mrad, including errors on alignments and track
extrapolation;
the local resolution of 13 (3.8) mrad is consistent with the design value.
The average number of detected photons per $\beta=1$ track was 16.1 (10.0) in
$\mu$-pair events.
In hadronic events, cuts to suppress spurious hits and cross-talk from
saturating hits gave an average of 12.8 (9.2) accepted hits.
The average reconstructed $\theta_c$ for $\beta=1$ tracks was 675 (58.6) mrad,
independent of position within the CRID and $\bar{\theta}_c^{liq}$ was constant
in time.
Time variations in $\bar{\theta}_c^{gas}$ of up to $\pm1.2$ mrad were tracked
with an online monitor and verified in the data.

Tracks were identified using a likelihood technique \cite{davea}. 
For each of the hypotheses $i=e,\mu,\pi,K$, p, a likelihood $L_i$ was
calculated based upon the number of detected photons and their measured
$\theta_c$, the expected number and $\bar{\theta}_c$, and a background term.
The background included overlapping Cherenkov radiation from other tracks in
the event and a constant term normalized to the number of hits in the relevant
TPC that were associated with no track.

\begin{figure}
 \hspace*{0.5cm}   
   \epsfxsize=6.2in
   \begin{center}\mbox{\epsffile{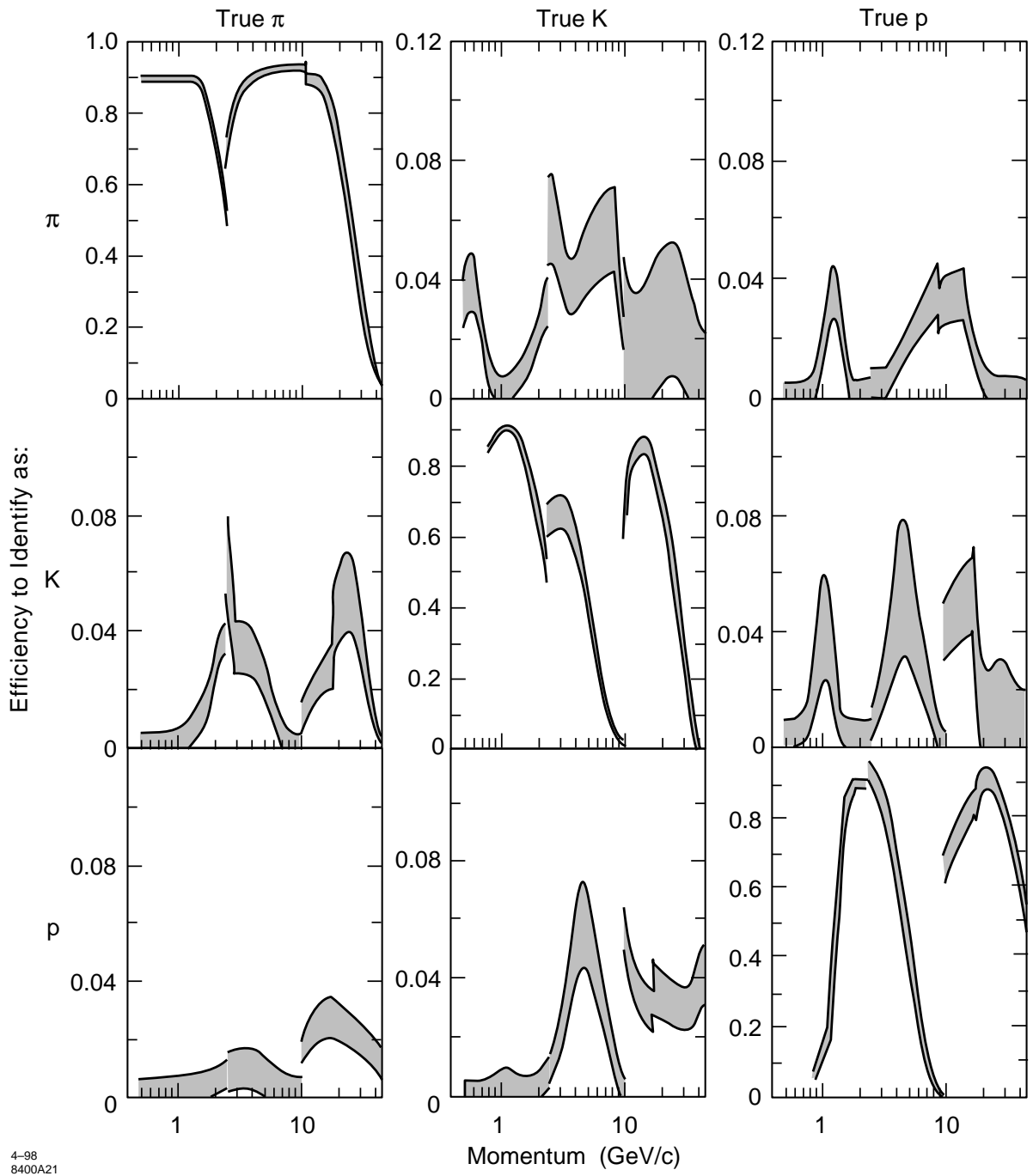}}\end{center}
  \caption{ 
 \label{effpar}
Calibrated identification efficiencies for tracks used in the charged hadron
analysis.
    }
\end{figure}

Cuts on differences between the logarithms of these likelihoods,
${\cal L}_i = \ln L_i$, are optimized for each analysis.
For example, in the analysis of charged hadrons (sec. 3) we considered only the
hypotheses $i=\pi$,$K$,p, and high purity was the primary consideration.
We therefore applied a tight set of CRID quality cuts \cite{bfp} (accepting
$\sim$60\% of the tracks with CRID information), and tracks with $p<2.5$
($p>2.5$) GeV/c were identified as species $j$ if ${\cal L}_j$ exceeded both
of the other log-likelihoods by at least 5 (3) units.
The matrix {\bf E} of identification efficiencies is shown in fig. \ref{effpar}.
The elements $E_{\pi j}$ and $E_{{\rm p} j}$ were determined from the data
using tracks from selected $K_s^0$, $\tau$ and $\Lambda^0$ decays.
The $E_{Kj}$ were related to the measured elements using a detailed
detector simulation.
The bands in fig.~\ref{effpar} encompass the systematic errors on the
efficiencies, determined from the statistics of the data test samples.
The discontinuities correspond to Cherenkov thresholds in the
gaseous radiator.
The identification efficiencies peak near or above 0.9 and the pion coverage
is continuous over $\sim$0.3--35 GeV/c.
There is a gap in the kaon-proton separation, $\sim$7--10 GeV/c, and
the proton coverage extends to the beam momentum.
Misidentification rates are typically less than 0.03, with peak values of
up to 0.06. 

\begin{figure}
 \hspace*{0.5cm}   
   \epsfxsize=6.6in
   \begin{center}\mbox{\epsffile{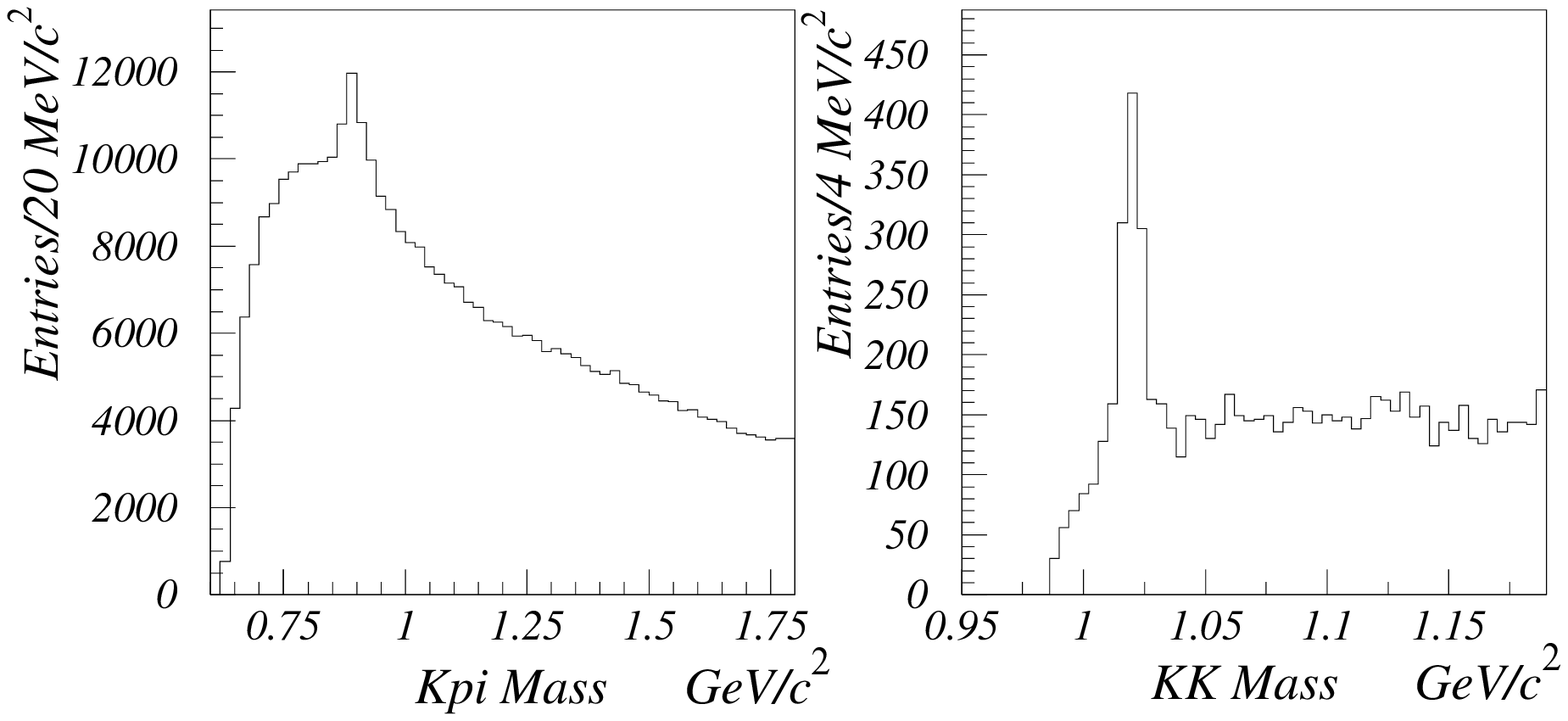}}\end{center}
  \caption{ 
 \label{kppeaks}
Distributions of invariant mass using loosely identified kaons (see text)
showing signals for the $K^{*0}$(890) and $\phi$(1020).
    }
\end{figure}

\begin{figure}
 \hspace*{0.5cm}   
   \epsfxsize=6.45in
   \begin{center}\mbox{\epsffile{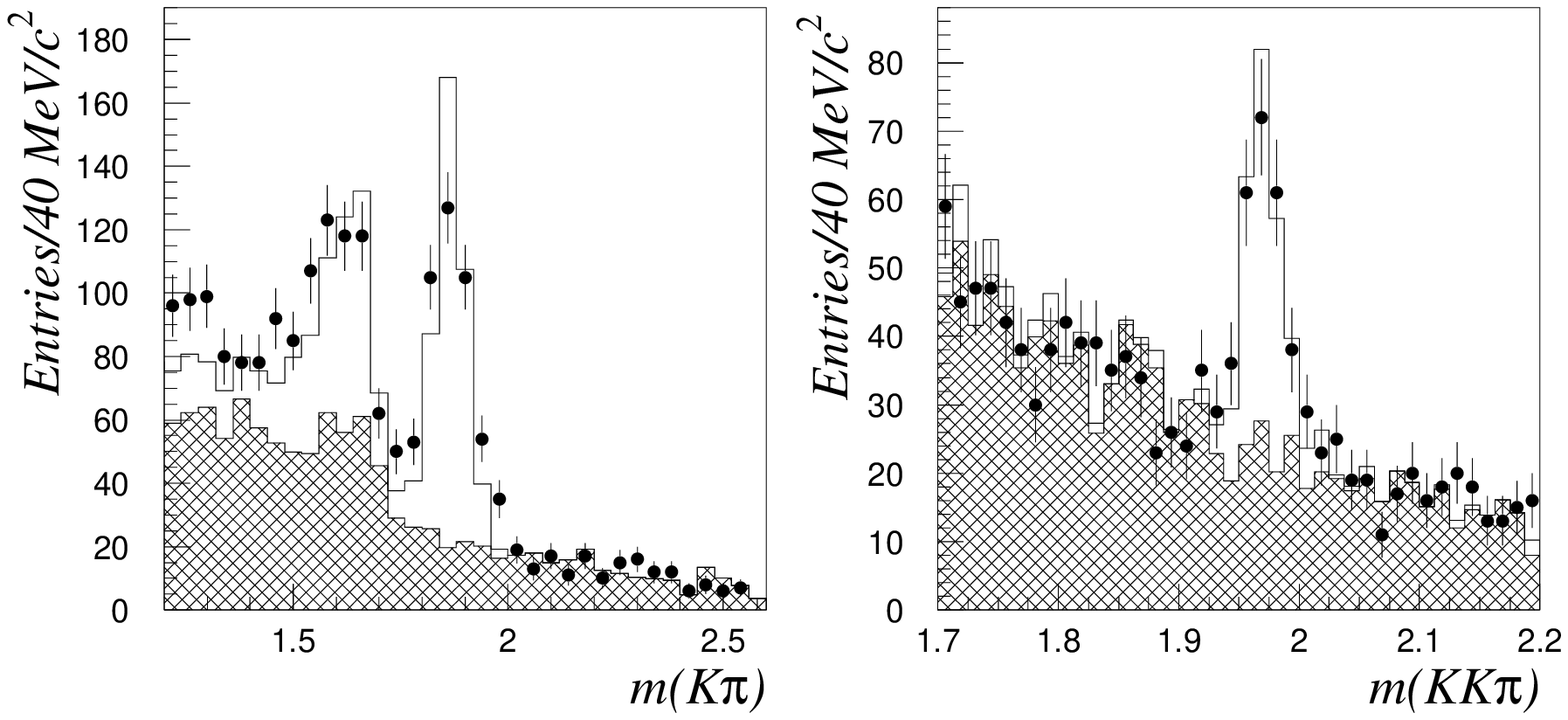}}\end{center}
  \caption{ 
 \label{dpeaks}
Invariant mass distributions for tracks forming secondary vertices (dots)
showing signals for the $D^{0}$, including the satellite peak, and
$D_s$ mesons.  The simulated (background) distributions are shown
as (hatched) histograms.
    }
\end{figure}

Many analyses required identifying $K^\pm$ with high efficiency and reasonable
purity.
A looser track selection was made ($\sim$95\% of CRID tracks)
and a moderate cut, typically ${\cal L}_K - {\cal L}_\pi > 3$, made against
pions and leptons, along with a loose cut against protons, typically
${\cal L}_K - {\cal L}_{\rm p} > -1$.
The efficiency for identifying true kaons is $\sim$70\% for
$1 < p < 30$ GeV/c;  the pion (proton) misidentification rate
depends strongly on momentum and can be as large as 12\% (70\%).
$K^\pm$ samples of 70--90\% purity are achieved.
The power of this loose identification for reconstructing strange and charmed
mesons is illustrated in figs. \ref{kppeaks} and \ref{dpeaks}.

The CRID has also been used in the identification of leptons.
Alone, it provides efficient $e$-$\pi$ separation for $p<4$ GeV/c, and
combined with the electromagnetic calorimeter gave improved $e^\pm$
identification for $p<8$ GeV/c.
For $e^\pm$ from $B$ hadron decays, adding the CRID information resulted in a
doubling of the efficiency of an optimized algorithm for a given purity.
For $\mu^\pm$, the CRID rejects $\pi^\pm$ for $2<p<4$ GeV/c,
and $K^\pm$, a substantial source of punchthrough, at all momenta.
Reoptimizing the muon selection including CRID information increased the
efficiency by $\sim$10\% at all $p$, and the purity by $\sim$40\% for
$2<p<4$ GeV/c and 5--30\% at higher momenta.

\newpage

\section{Hadronization Physics}

Inclusive properties of the charged tracks and photons in jets have been studied
extensively in $e^+e^-$ annihilations.
Studies of specific identified particles at lower energies had
low statistics and incomplete momentum coverage, but were able to observe
the production of baryons, vector mesons and strange mesons and baryons,
and to study mechanisms for strangeness and baryon number conservation through
correlations.
The large samples at the $Z^0$ have allowed much more detailed studies
\cite{bohrer}, including the recent observations of tensor mesons and orbitally
excited baryons.

\begin{figure}
\vspace{-1.cm}
 \hspace*{0.5cm}   
   \epsfxsize=4.2in
   \begin{center}\mbox{\epsffile{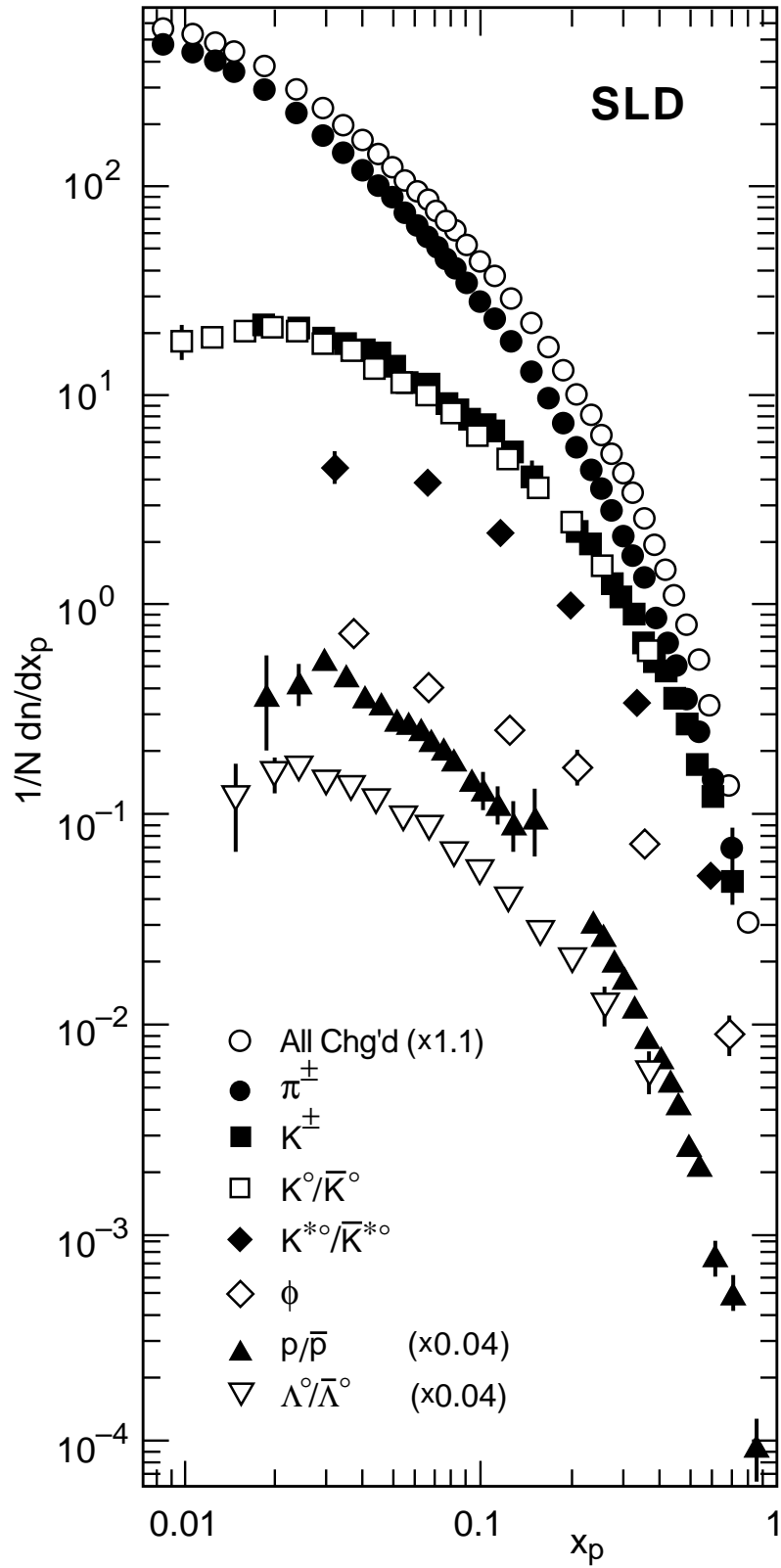}}\end{center}
  \caption{ 
 \label{xsall}
Differential production cross sections per hadronic $Z^0$ decay per unit
$x_p$ for all charged particles and seven identified hadron species.
    }
\end{figure} 

We have studied the production of seven identified particle species in
hadronic $Z^0$ decays \cite{bfp}.  Charged $\pi^\pm$, $K^\pm$ and
p/$\bar{\rm p}$ identified as described in sec. 2 were counted as a
function of momentum and these counts unfolded using the inverse of the
identification efficiency matrix (fig. \ref{effpar}) to yield production 
cross sections as a function of momentum.
The neutral strange vector mesons $K^{*0}$ and $\phi$ were reconstructed in
their $K^+\pi^-$ and $K^+K^-$ modes, respectively, using the loose kaon
selection (sec. 2).
Production cross sections were extracted from fits to the invariant mass
distributions (see fig. \ref{kppeaks}).
These cross sections, along with similar measurements for $K^0$ and
$\Lambda^0$/$\bar{\Lambda}^0$ \cite{bfp}, are shown in fig.~\ref{xsall}.

These measurements cover a wide momentum range with good precision.
Similar inclusive measurements have been made \cite{bohrer} at the $Z^0$ by
DELPHI using RICH particle identification, and by ALEPH and OPAL
using d$E$/d$x$.
The measurements are consistent, have comparable precision and, between the two
methods, cover the entire momentum range.
The clean samples available from the CRID compensate for the much higher
statistics at LEP, and also allow smaller systematic errors in some cases, most
notably the $K^{*0}$, for which the background is high and contains large
contributions from reflections of resonances decaying into $\pi^+\pi^-$.

\begin{figure}
\vspace{-1.cm}
 \hspace*{0.5cm}   
   \epsfxsize=4.0in
   \begin{center}\mbox{\epsffile{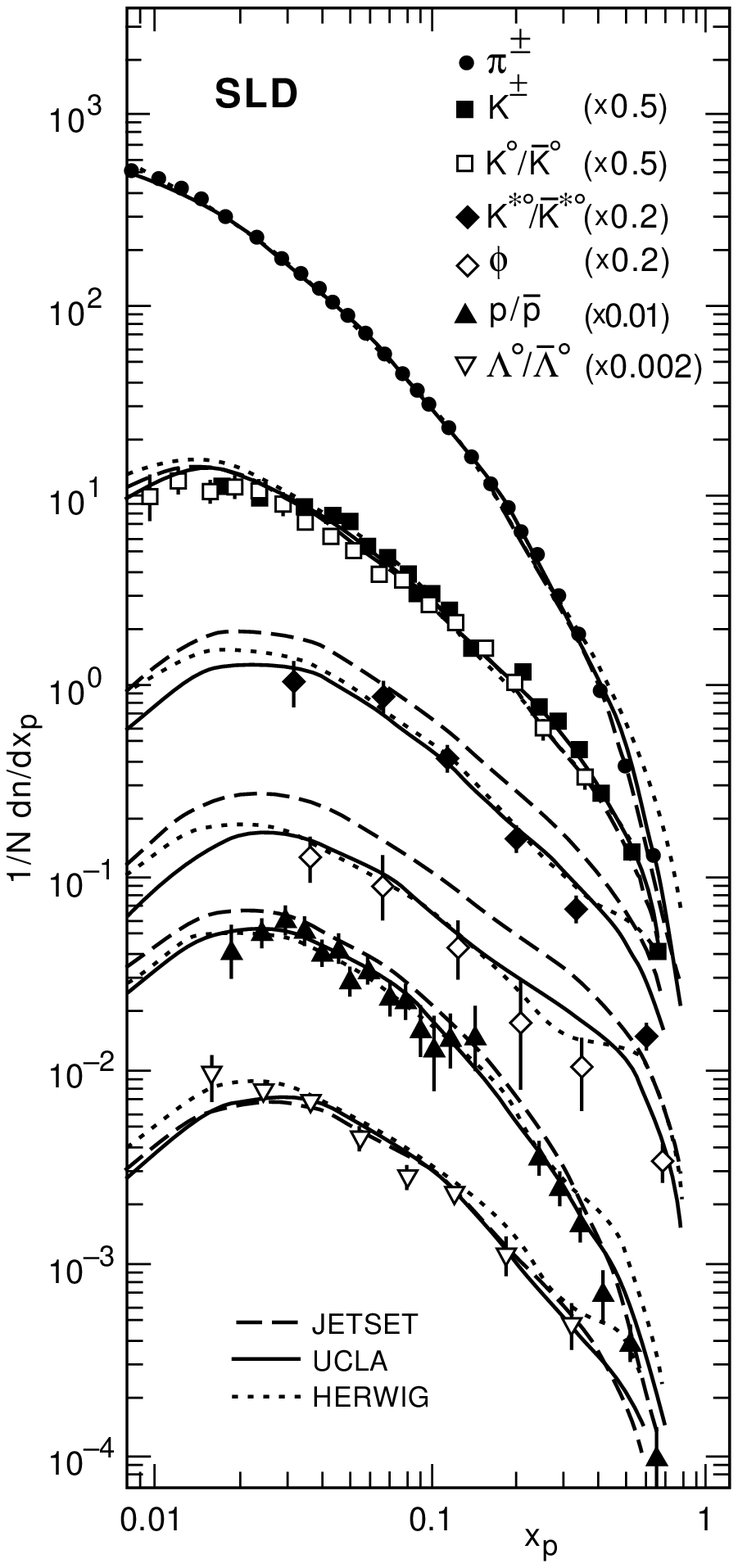}}\end{center}
  \caption{ 
 \label{xsudsmc}
Differential production cross sections per light-flavor hadronic $Z^0$ decay,
compared with the predictions of three hadronization models.
    }
\end{figure} 

The CRID efficiency and purity are especially useful when the data are divided
into smaller samples and additional levels of unfolding are required.
We divided the hadronic events into $b$-, $c$- and light-flavor
($u$,$d$,$s$) samples \cite{bfp}, repeated the above analyses on
each, and unfolded the results to yield production cross sections in
these three flavor categories.
Substantial differences are observed, as expected from the known production and
decay properties of the leading heavy hadrons.
The results for the light-flavor sample are shown in fig. \ref{xsudsmc}, where
coverage and precision comparable to that of the flavor-inclusive sample are
evident.
This measurement provides a more pure way of looking at hadronization at a
fundamental level.
We have found these results to be consistent with the limited predictions of
perturbative QCD \cite{bfp}, and compared them with the predictions of three
hadronization models (see fig. \ref{xsudsmc}).
All describe the data qualitatively; differences in detail include:
all models are high for low-momentum kaons;
the JETSET model is high for the vector mesons and protons at all momenta;
the HERWIG model and, to a lesser extent, the UCLA model show excess structure
at high momentum for all particle species;
no model is able to reproduce the $\sim$10\% difference between $K^\pm$ and
$K^0$ production.
These discrepancies have been observed previously \cite{bohrer}, however our
measurement
demonstrates unambiguously that they are in the hadronization part of the model
and not in the simulation of heavy hadron production and decay.
We have found additional minor discrepancies in the heavy flavor
events \cite{bfp}.

\begin{figure}
\vspace{-1.cm}
 \hspace*{0.5cm}   
   \epsfxsize=3.4in
   \begin{center}\mbox{\epsffile{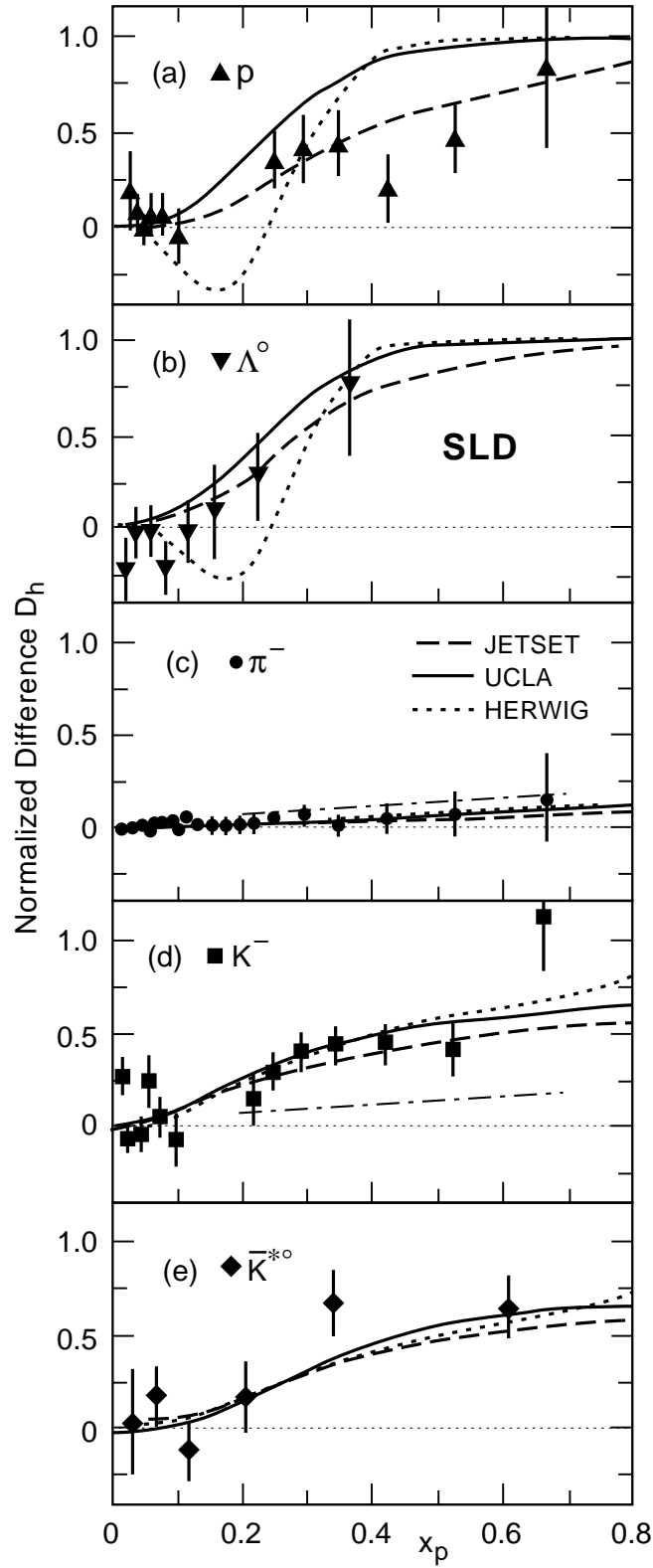}}\end{center}
  \caption{ 
 \label{ndall}
Normalized differences between hadron and antihadron production in $uds$
jets; the predictions of the three fragmentation models.
    }
\end{figure} 

The events in the light flavor sample can be divided further into quark
hemispheres and antiquark hemispheres using the electron beam polarization,
providing a unique study of leading particle effects \cite{lpprl,bfp}.
For events with thrust axis $|\cos\theta|>0.2$, the forward (backward) thrust
hemisphere is tagged as the quark jet if the beam is left-(right-)polarized,
and the opposite hemisphere tagged as the antiquark jet.
The SM predicts a quark purity of 73\%.
The set of hadrons in quark jets plus their respective antihadrons in antiquark
jets were analyzed to yield cross sections $\sigma_h$ for hadrons in light
quark jets.
Similarly, the remaining hadrons yielded antihadron cross sections
$\sigma_{\bar{h}}$.
The corrected normalized production differences
$D_h = (\sigma_h - \sigma_{\bar{h}}) / (\sigma_h + \sigma_{\bar{h}})$ are
shown in fig. \ref{ndall}.

At low momentum, hadron and antihadron production are consistent.
At higher momentum there is an excess of baryons over antibaryons, as
expected from leading baryon production (a baryon contains valence quarks,
not antiquarks).
The large excess of pseudoscalar and vector antikaons over kaons at high
momentum is evidence both for leading kaon production, and for the dominance of
$s\bar{s}$ events in producing leading kaons.
No significant leading particle signature is visible for pions.

CRID type particle identification greatly enhances studies of pairs of hadrons
in the same event.
We have analyzed \cite{correls} correlations in rapidity
$y=0.5\ln((E+p_{\parallel})/(E-p_{\parallel}))$, where $E$ ($p_{\parallel}$)
is the energy (momentum projection onto the thrust axis) of the hadron, between
pairs of identified $\pi^\pm$, $K^\pm$ and p/$\bar{\rm p}$ in light-flavor
events.
We compared the distribution of $\Delta y=|y_1-y_2|$ for identified $K^+K^-$
pairs with that for $K^+K^+$ and $K^-K^-$ pairs.
The latter are expected to be uncorrelated, and the difference between the two
distributions illuminates strangeness production in the hadronization process.
We observe \cite{correls} a large difference at low values of $\Delta y$; this
`short range' correlation indicates that the conservation of strangeness is
`local', that is, a strange and an antistrange particle are produced close to
each other in the phase space of the jet.
Similar effects for p$\bar{\rm p}$ and $\pi^+\pi^-$ pairs indicate local
conservation of baryon number and isospin, respectively.
Such effects have been observed previously, however the CRID has allowed the
study of the shape and range of the correlations in detail and at many momenta,
in particular we have verified the scale-invariance of the range.
We have also observed short-range correlations between opposite-charge $\pi K$,
$\pi$p and $K$p pairs; they are relatively weak, and high purity is required to
separate them from the large $\pi\pi$ background.
They suggest charge-ordering of all particle types along the $q\bar{q}$ axis
and provide new tests of fragmentation models.

\begin{figure}
   \epsfxsize=6.9in
   \begin{center}\mbox{\epsffile{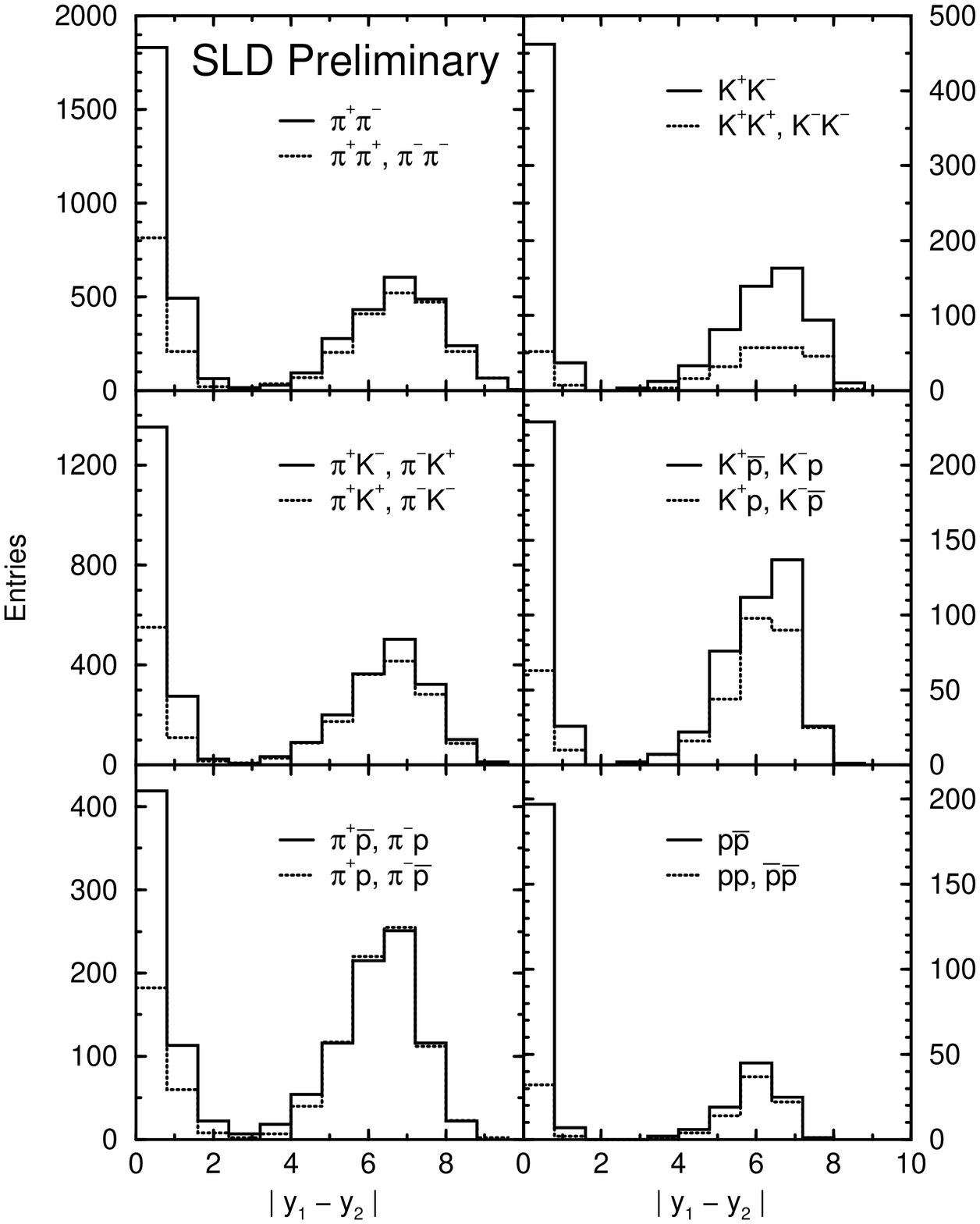}}\end{center}
  \caption{ 
 \baselineskip=14pt
 \label{fdrap9}
$|\Delta y|$ distributions for opposite- (histograms) and
same-charge (dashed histograms) pairs in which both tracks have $p>9$ GeV/c.
}
\end{figure} 

We also expect correlations at long range due to leading particles, e.g.
an $s\bar{s}$ event may have a leading $K^-$ in the $s$ jet and a leading $K^+$
in the $\bar{s}$ jet that have a large $\Delta y$.
We have studied pairs of identified hadrons that both have $p>9$ GeV/c.
Their $\Delta y$ distributions are shown in fig. \ref{fdrap9};  the $p$ cut
separates each distribution into two parts, one ($\Delta y<2$) comprising pairs
in the same jet and the other ($\Delta y>3$) comprising pairs in opposite jets
of the event.
Strong $K^+K^-$ correlations are seen at both short and long range, as expected.
For baryon and pion pairs, any long-range correlation will be diluted by the
short-range correlation -- a high-momentum leading baryon will always be
accompanied by a subleading antibaryon, also with high momentum.
We do not observe a long-range correlation for p$\bar{\rm p}$ pairs, however we
do observe a significant correlation for $\pi^+\pi^-$ pairs, providing direct
evidence for leading pion production.
There are also significant correlations for $\pi K$ and $K$p pairs, but not for
$\pi$p pairs.
These cross terms provide new information on leading particle production in
jets of different flavors, which will eventually allow the use of high momentum
identified particles to separate $u\bar{u}$, $d\bar{d}$ and $s\bar{s}$ events
from each other.

Using the beam polarization to select the quark hemisphere
in each event, we have performed a new study \cite{correls} of rapidities
signed such that $y>0$ ($y<0$) corresponds to the (anti)quark
direction.
A pair of identified hadrons can then be ordered, for example by charge to
form the ordered rapidity difference $\Delta y^{+-} = y_+ - y_-$.
A positive value of $\Delta y^{+-}$ indicates that the positively charged
hadron is more in the direction of the primary quark than the negatively
charged hadron.
The distribution of $\Delta y^{+-}$ can be studied in terms of the
difference between its positive and negative sides.
We observe a large difference for $K^+K^-$ pairs at long range due to
leading kaon production in $s\bar{s}$ events.
A significant difference at short range for p$\bar{\rm p}$ pairs at all $p$ is
direct evidence that the proton in a correlated p$\bar{\rm p}$ pair prefers the
quark direction over the antiquark direction.

\section{Quark Flavor Tagging and Electroweak Physics}

The identification of the flavor of the quark that initiated a hadronic jet is
required for a wide variety of physics.
The development of precision vertex detectors has enabled the pure and efficient
tagging of $b$/$\bar{b}$ and $c$/$\bar{c}$ jets, leading to a number of precise
measurements in fixed-target and $e^+e^-$ annihilation experiments
and the understanding of top quark production at hadron colliders.

For many measurements it is also necessary to distinguish $b$ from $\bar{b}$
or $c$ from $\bar{c}$ jets.
The use of charged kaons is envisioned for this purpose in several future $B$
physics experiments, and has recently been pioneered by SLD.
We present examples of its use for both $b$ and $c$ physics, which rely on the
CRID for high efficiency and purity that is measurable from the data.

The identification of light flavor jets is a field in its infancy.
The three light flavors can be separated from $b$/$\bar{b}$ and $c$/$\bar{c}$
jets by the absence of a secondary vertex in the jet;
the only known way to distinguish them from each
other is by identifying the leading particle in the jet cleanly.
Little experimental information on leading particle production in light flavor
jets exists (much of which appears in sec.~3 above), making the measurement of
tagging purities and analyzing powers from the data essential.
We present a measurement of $A_s$ and discuss prospects for light-flavor tagging
in general.

\subsection{$B^0$-$\bar{B}^0$ Mixing}

To measure the time dependence of $B^0$-$\bar{B}^0$ mixing, one must tag
neutral $B$ hadrons and determine their flavor ($B$ or $\bar{B}$) at both
production and decay time.
We first selected a sample of high-mass secondary vertices \cite{topo}
(98\% $B$-purity, $\sim$40\% $B^0_d$) and reconstructed the proper decay time
$\tau$ of each.
The flavor at production was determined by a combination of the beam
polarization and the charges of the tracks, identified $K^\pm$, $e^\pm$ and
$\mu^\pm$ in the opposite hemisphere, with a correct-sign
fraction of 88\%.
The flavor at decay time was determined from the charge of any
identified $K^\pm$ attached to the secondary vertex, with a correct-sign
probability $P_{corr}=77$\%.

\begin{figure}
 \hspace*{0.5cm}   
   \epsfxsize=4.9in
   \begin{center}\mbox{\epsffile{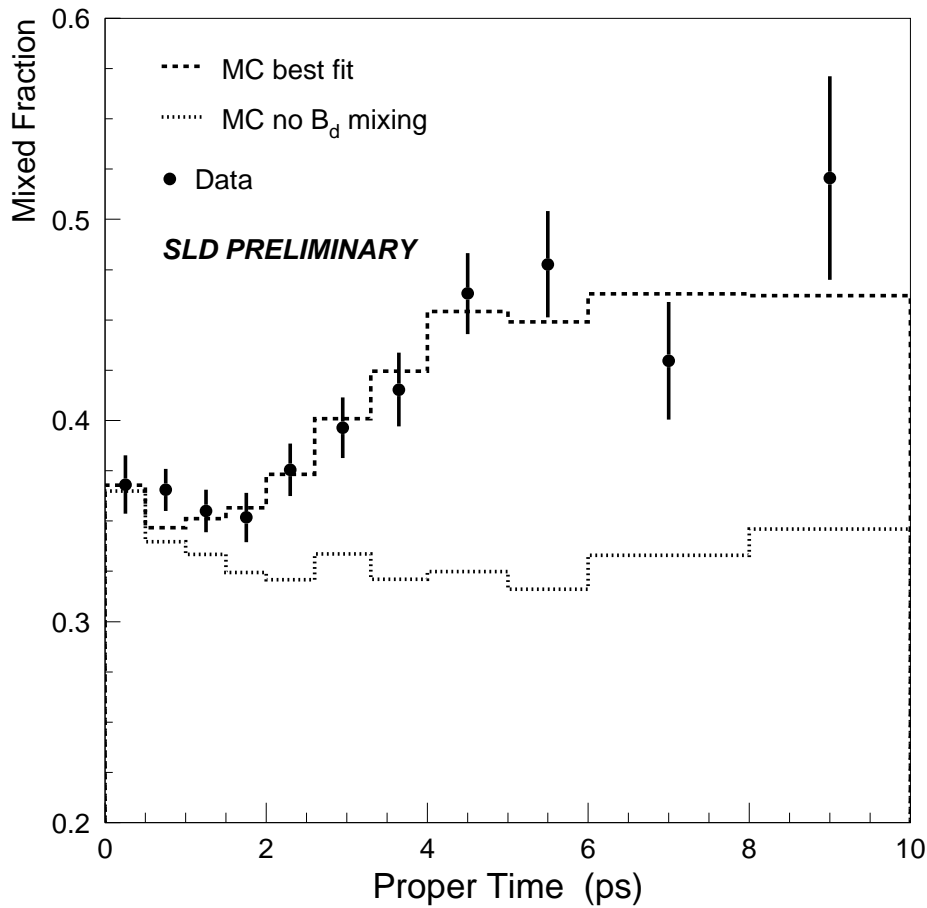}}\end{center}
  \caption{ 
 \label{bmix}
Mixed fraction as a function of proper decay time for tagged $B$ decay vertices.
    }
\end{figure} 

The fraction of events classified as mixed (different flavors at production and
decay) is shown as a function of $\tau$ in fig. \ref{bmix}.
A clear increase with time is evident, which is the signal for
$B^0_d$-$\bar{B}^0_d$ mixing.
A fit to the data yielded a 6\% measurement of the mass difference $\Delta m_d$.
There are several other measurements of $\Delta m_d$ on the market with
similar precision, including three from SLD, however this is the
only one using the $K^\pm$ tag, providing valuable complementarity.

There is considerable interest in $B^0_s$-$\bar{B}^0_s$ mixing, for which there
are currently only lower limits on the frequency.
Our $K^\pm$ charge tag is insensitive to the $B_s$ flavor, since the $B^0_s$
decay contains both a $K$ and $\bar{K}$ meson, which is a feature of the
above $B^0_d$-$\bar{B}^0_d$ mixing measurement.
However, the measurement of $B^0_s$-$\bar{B}^0_s$ mixing requires a strongly
enriched $B_s$ sample, for which the CRID is quite useful
(see fig. \ref{dpeaks}).
Also, the subleading $K^-$ produced in association with a $\bar{B}_s^0$ meson
can be used as an initial state tag, as demonstrated in a $B_s$ production
measurement by DELPHI \cite{marco}.

\subsection{Heavy Flavor Asymmetries}

In order to measure the asymmetric coupling $A_q$ of the $Z^0$ to quarks of
flavor $q$ it
is necessary to identify the event flavor as $Z^0 \rightarrow q\bar{q}$ and to
determine the polar angle $\theta_q$ of the primary $q$.  The event thrust axis
provides a good estimate of $|\cos\theta_q|$, however it must still be
determined which thrust hemisphere contains the initial $q$ and which the
$\bar{q}$.

In the case of the heaviest flavors accessible at the $Z^0$, $q=b,c$, the event
flavor can be determined by the presence of a secondary vertex of high or low
invariant mass, respectively.
We selected samples of 96\% $b\bar{b}$ purity and 69\% $c\bar{c}$ purity,
respectively, and measured these purities in the data.
We have used four methods so far to determine the primary $b$ or $c$ direction,
summarized in table \ref{tftags}.

\begin{table}
\begin{center}
\begin{tabular}{|l|l|c|c|c|c|c|c|}
\hline
&&&& \multicolumn{2}{c|}{  } & \multicolumn{2}{c|}{  } \\[-0.3cm]
Flavor Tag & Rel. & Correct Sign  & Corr. Frac. &
                    \multicolumn{2}{c|}{Error on $A_b$} &
                    \multicolumn{2}{c|}{Error on $A_c$} \\
Method     & Eff. & Fraction b/c  & Measmt.     & Now & Final
                                                & Now & Final \\[0.1cm]
\hline 
&&&&&&&\\[-0.3cm]
Evt. Chg. $Q$  & 1.0  &  $\sim$0.6, $|Q|$-dep.&  model-dep.  &
               0.040  &  0.035      &   --    &       --     \\
Id'd $K^\pm$   & 0.4  &  $\sim$0.75 / 0.90    &  clean       &
               0.138  &  0.040      & 0.050   &  0.036       \\
Id'd lepton    & 0.1  &  $\sim$0.90 / 0.95    &  stat. ltd.  &
               0.069  &  0.044      & 0.110   &  0.070       \\
Rec'd $D^{(*)}$& 0.01 &  $\sim$0.85 / 1.00    &  stat. ltd.  &
                 --   &  --         & 0.072   &  0.044       \\[0.1cm]
\hline 
\end{tabular}
\caption{\label{tftags} 
Summary of SLD heavy flavor asymmetry measurements}
\end{center}
\end{table}

A standard method is to use identified $e^\pm$ and $\mu^\pm$, which were
assigned to $B$ decays, $D$ decays, or cascade $B\rightarrow D\rightarrow l$
decays based on the vertex mass and the lepton $p$ and $p_t$ with respect to
the vertex flight direction.
An $e^-$ or $\mu^-$ tags a $\bar{B}$ or $\bar{D}$ meson, and a likelihood fit
was used to extract $A_b$ and $A_c$ simultaneously.
This method is competitive due to the CRID-assisted $e^\pm$ and $\mu^\pm$
identification (sec. 2) and the high $P_{corr}$.
However the efficiency is low and the sources of leptons overlap
substantially, making it difficult to measure $P_{corr}$ in the data.

We have measured $A_c$ using $D$ and $D^*$ mesons reconstructed in several
decay modes.
The CRID is essential for the $D^0\rightarrow K^-\pi^+$ (with no $D^*$
requirements) and $D_s\rightarrow K^+K^-\pi^+$ modes (see fig. \ref{dpeaks}).
This method has essentially unit $P_{corr}$, but suffers from relatively low
statistics.

A standard method for $A_b$ is the momentum-weighted event charge
$Q$ \cite{abjetc}, which 
%$Q=\Sigma_jq_j {\rm sgn}(\vec{p}_j\cdot\hat{t})|\vec{p}_j\cdot\hat{t}|^{0.5}$,
%where $q_j$ ($\vec{p}_j$) is the charge
%(momentum) of the $j^{th}$ track in the event and $\hat{t}$ is the thrust axis.
%The axis was signed such that $Q<0$ and used to estimate $\cos\theta_b$.
%This method is able to give a sign to every tagged $b\bar{b}$ event, but
gives a sign to every tagged $b\bar{b}$ event, but
$P_{corr}$ is low, averaging $\sim$60\%, and depends on $|Q|$.
$P_{corr}$ can be measured in the data under some model-dependent assumptions,
but the method will reach a systematic limit of a few percent.

\begin{figure}
 \hspace*{0.5cm}   
   \epsfxsize=6.3in
   \begin{center}\mbox{\epsffile{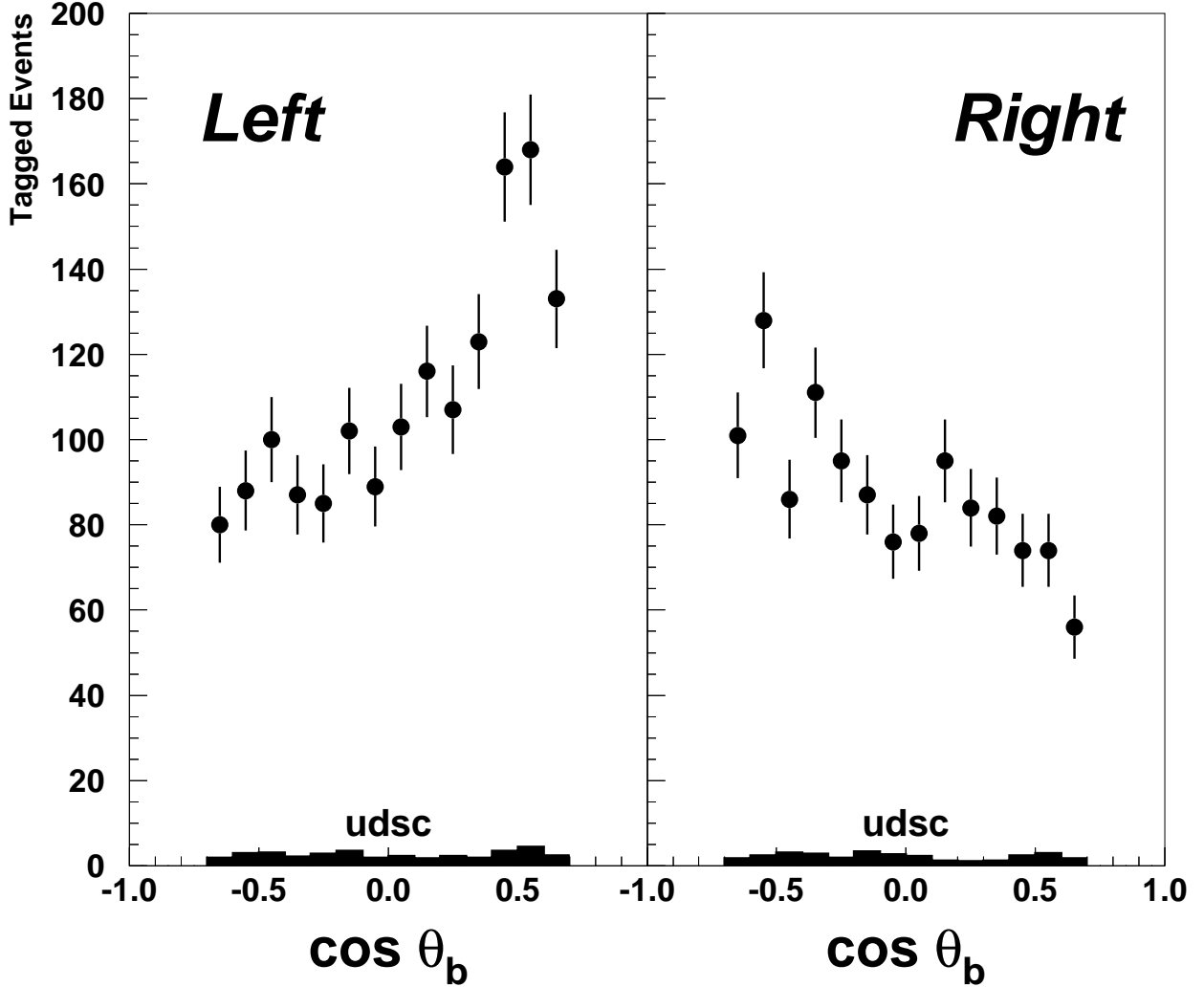}}\end{center}
  \caption{ 
 \label{cthb}
Distributions of the $b$-quark polar angle for events produced with negatively
(left) and positively (right) polarized electron beams.
    }
\end{figure} 

We have recently demonstrated the use of identified $K^\pm$ to measure both
$A_b$ \cite{abkaon} and $A_c$ \cite{ackaon}.
A $K^\pm$ was identified in $\sim$40\% of the tagged $B$ hadron vertices with
$P_{corr}=73$\%, a smaller value than in sec.~4.1 since $B^0$-$\bar{B}^0$
mixing is now a dilution.
This method provides a nice balance between efficiency and purity, yields the
$\cos\theta_b$ distributions in fig. \ref{cthb}, and $P_{corr}$ can be measured
unambigusouly in the data.
The precision of the $P_{corr}$ measurement depends on the square of the
efficiency and is also quite sensitive to background, so that CRID type $K^\pm$
identification is essential.
Currently this measurement is statistics limited, as we have analyzed less than
one-sixth of our data sample; 
we expect to measure $P_{corr}$ to $\pm2$\%, a
valuable result in itself for future $B$ physics experiments, and to achieve
one of the best precisions on $A_b$.
Furthermore, this is a unique method complementary to those used so far, which
give a world average value of $A_b$ that differs from the SM prediction by
3$\sigma$.

\begin{figure}
 \hspace*{0.5cm}   
   \epsfxsize=7.65in
   \begin{center}\mbox{\epsffile{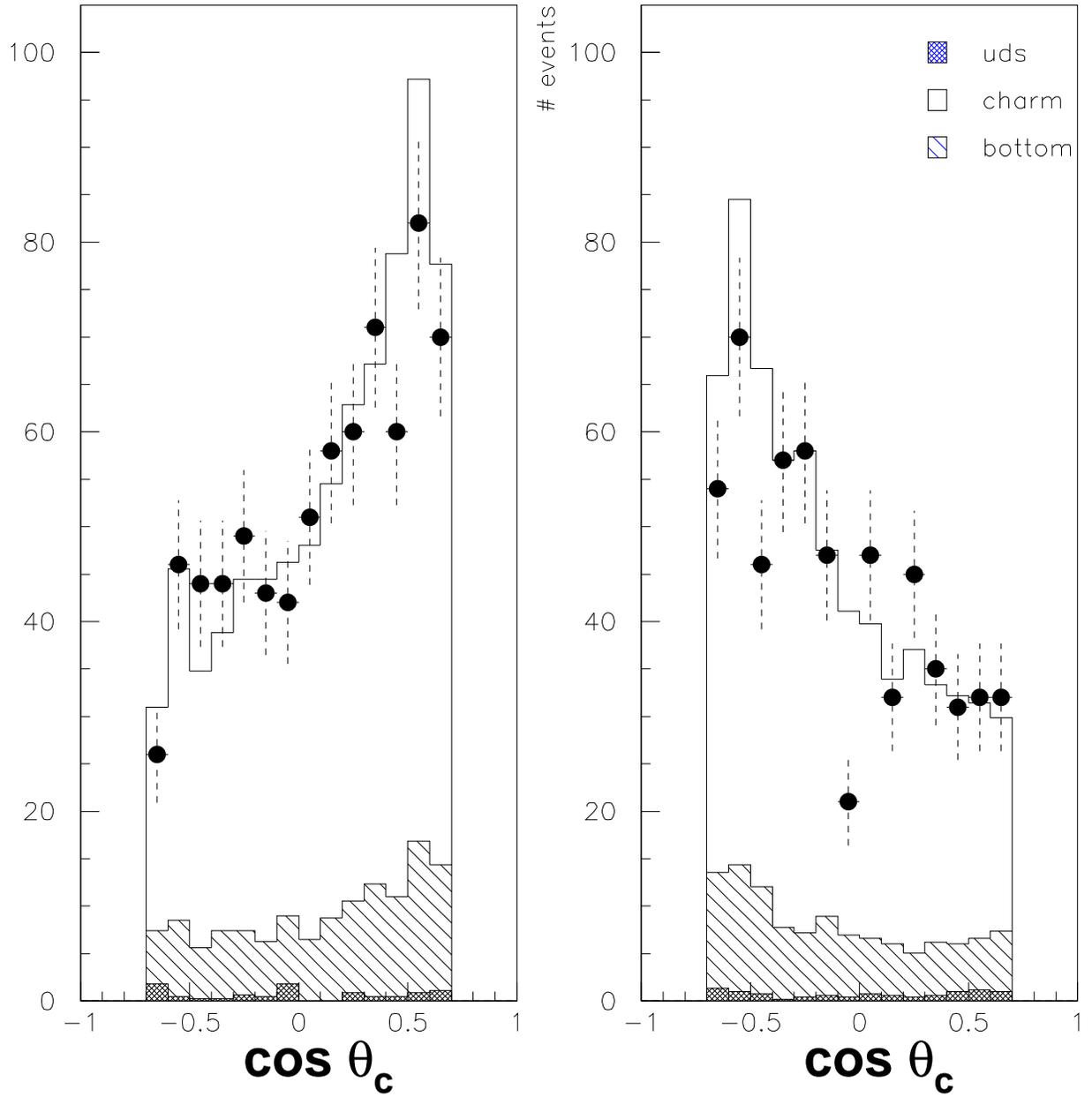}}\end{center}
  \caption{ 
 \label{cthc}
Distributions of the $c$-quark polar angle for left- and right-polarized
beams.
    }
\end{figure} 

The situation is even better for charm.
A $K^\pm$ was found in $\sim$40\% of low mass vertices with zero net charge,
and gave $P_{corr}>90$\%.
In charged vertices the vertex charge was combined with that of any identified
$K^\pm$ to give an average $P_{corr}=91$\% for all charm vertices,
and the $\cos\theta_c$ distributions shown in fig. \ref{cthc}.
This $P_{corr}$ can be estimated reliably from known charmed hadron branching
ratios, and with one-half of our data analyzed, we have the world's best
measurement of $A_c$.
A measurement of $P_{corr}$ from the data is feasible and will be
required to reach 1\% precision.

\subsection{Light Flavor Asymmetries}

In contrast to the situation with leptons or heavy flavors, there are published
measurements of $Z^0$ couplings to light-flavor quarks ($u$, $d$ and $s$)
only from DELPHI \cite{asdelphi} and OPAL \cite{alopal}, with rather poor
precision.
The challenge is to separate these flavors not only from the heavy flavors but
also from each other.
Leading particles at high momentum can be used to determine the event flavor,
and, if they carry the appropriate quantum number, the direction of the quark.
However, the lack of experimental measurements of leading particle effects for
these flavors leads to the choice of relying on a hadronization model to predict
sample purities and $P_{corr}$, or trying to measure these in the data.

We have taken the approach of applying hard flavor selection cuts to reduce
backgrounds, increase the $P_{corr}$ and therefore reduce the associated
systematic errors.
A hadronization model was used to predict these, but key quantities were
measured in the data, the predictions adjusted accordingly, and the data
statistics used to estimate the systematic errors.

\begin{figure}
 \hspace*{0.5cm}   
   \epsfxsize=7.9in
   \begin{center}\mbox{\epsffile{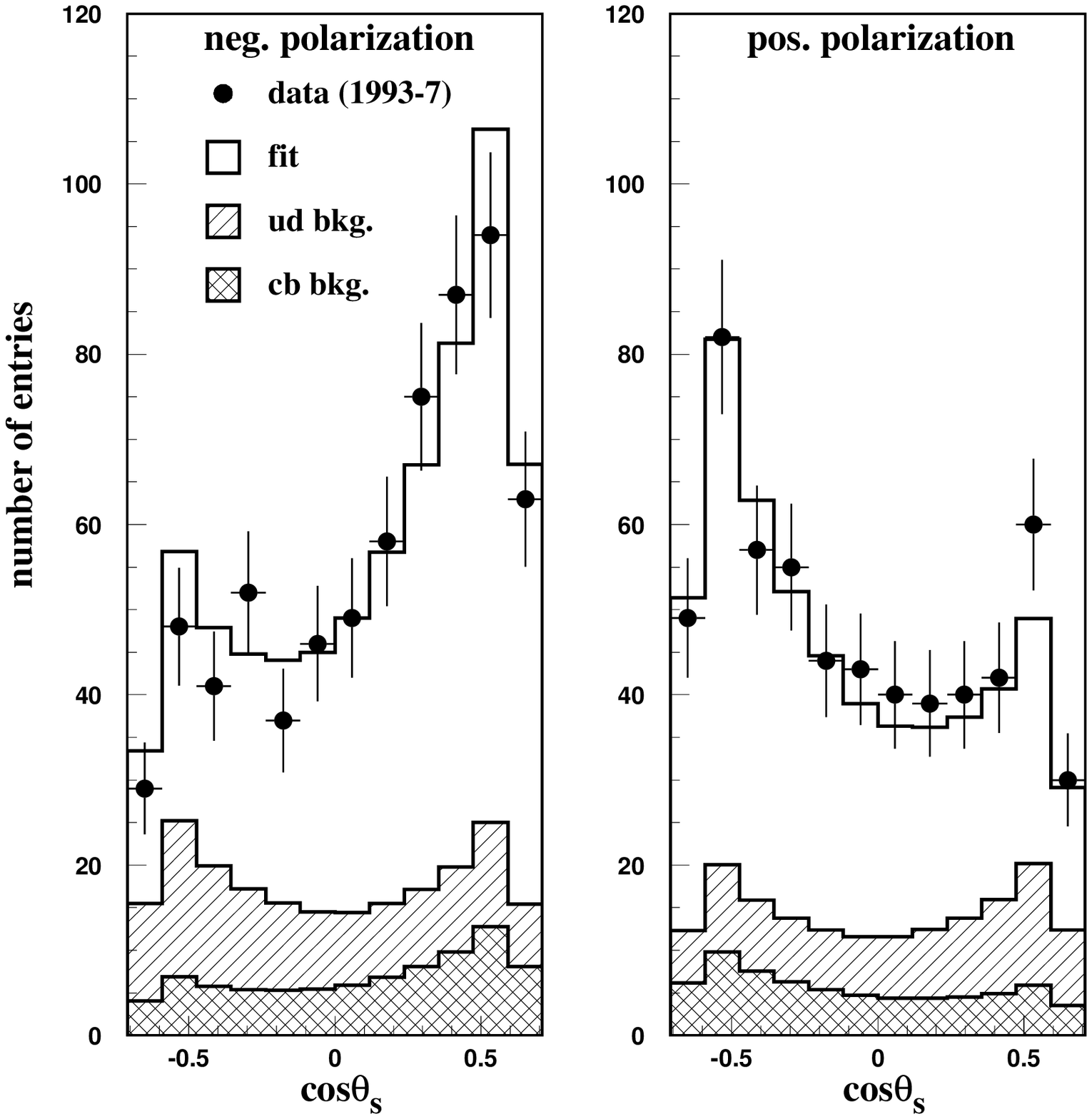}}\end{center}
  \caption{ 
 \label{cths}
Distributions of the $s$-quark polar angle for left- and
right-polarized beams.
    }
\end{figure} 

To measure $A_s$ \cite{askaon}, we used the light flavor sample and tagged $s$
($\bar{s}$)
hemispheres using identified $K^-$ ($K^+$) with $p>9$ GeV/c and reconstructed
$\Lambda^0\rightarrow {\rm p}\pi^-$
($\bar{\Lambda}^0\rightarrow \bar{\rm p}\pi^+$) with $p>5$ GeV/c and the
(anti)proton identified in the CRID.
Requiring either an $s$-tag in one hemisphere and an $\bar{s}$-tag in the other,
or an $s$- or $\bar{s}$-tag in one hemisphere and a reconstructed
$K_s^0 \rightarrow \pi^+\pi^-$ with $p>5$ GeV/c in the other, yielded an event
sample with 69\% $s\bar{s}$ purity and an average $P_{corr}=82$\%.
The distributions of $\cos\theta_s$ are shown in fig.~\ref{cths}.

The heavy-flavor background is substantial but understood, and the
associated systematic errors are small.
$P_{corr}$ for $s\bar{s}$ events was measured in the data by counting
hemispheres in which we identified three $K^\pm$ or $K^0_s$, since hemispheres
with three true kaons are the dominant source of wrong signs.
Similarly, the level and asymmetry of the $u\bar{u}$/$d\bar{d}$ background were
measured by counting hemispheres with two identified kaons, and events with an
$s$-tag (or $\bar{s}$-tag) in both hemispheres.
The simulation was used to relate these counts to the relevant quantities,
and was checked against our measured $KK$ correlations (sec.~3).

The CRID was essential in this analysis, yielding both a clean $s\bar{s}$
sample and enough 2- and 3-kaon hemispheres to measure the $P_{corr}$
and background.
A fit gave $A_s=0.82\pm 0.10\pm 0.07$ (Preliminary), which is consistent with
both the SM and the world average $A_b$, and is already a useful test of
down-type universality.
It is consistent with previous measurements and with a new measurement
from DELPHI \cite{newdelphi}.
We expect a total relative uncertainty of $\sim$9\% when our full data sample is
analyzed, and with another data run or further input from the LEP experiments
we might hope for a world average with a $\sim$5\% uncertainty

An open question is the extent to which we can measure the other two light
flavor asymmetries, $A_u$ and $A_d$.
The two flavors are expected to have similar production of leading pions
and protons, but perhaps to  differ in leading strange particle production.
Here it is essential to measure the purities and $P_{corr}$ in the data,
as has been done by OPAL \cite{alopal}.
We are pursuing this topic and expect $\sim$20\% measurements based
on the statitstics of our long-range correlation data (fig. \ref{fdrap9}).

\section{Conclusions}

We have presented a number of recent results from the SLD collaboration that use
the CRID for charged particle identification.
The performance of both the liquid and gaseous radiator systems has reached
essentially the design parameters in terms of
Cherenkov photon yield, angular resolution and hadron identification.
CRID and calorimeter information have been combined to enhance the
identification of electrons and muons.
A wide variety of physics has been made possible or improved by the use of the
CRID, in such diverse areas as hadronic jet structure, $B$-hadron decays and
precision electroweak physics.

We have made precise studies of the production of $\pi^\pm$, $K^\pm$,
p/$\bar{\rm p}$, $K^{*0}$ and $\phi$ in inclusive hadronic events,
complementing measurements made with other particle identification techniques.
We have also made new studies of events of different flavors and
of leading particles.
We have made the first precise studies of short-range, long-range and ordered
correlations between pairs of identified $\pi^\pm$, $K^\pm$ and p/$\bar{\rm p}$
in light-flavor events.
This set of results greatly enhances our understanding of jet formation.

The complete momentum coverage of $K^\pm$ from $B$ decays, combined with
precision vertexing, has allowed a number of studies of $B$ hadron decays 
that are difficult at existing $\Upsilon$(4S) detectors.
We have studied the endpoint of the $K^\pm$ spectrum, finding no
evidence for $b\rightarrow sg$ transitions.
We have compared $K^\pm$ production from the up- and downstream vertices
in inclusive $B$ decays, and extracted a new measurement of the fraction of
doubly charmed $B$ decays.
Such results demonstrate the power of
full momentum coverage, and bode well for future
dedicated $B$ physics experiments that include a RICH.

Flavor tagging of hadronic jets is an important recent advance in elementary
particle physics.
We have pioneered the use of $K^\pm$ to distinguish $B$ from $\bar{B}$ hadrons
and $b$ from $\bar{b}$ jets inclusively, envisioned as an efficient and
powerful tool at future dedicated $B$ physics experiments.
Our measurements of $\Delta m_d$ and $A_b$ are competitive, with
uncertainties dominated by the unknown analyzing power of this technique.
When all our data are analyzed we expect to measure this analyzing power to
$\pm$2\% and to obtain measurements of $\Delta m_d$, $\Delta m_s$ and $A_b$
that are among the world's best and complement existing techniques.
This inclusive method is also effective for charm; we already have the world's
best measurement of $A_c$, and this technique should find applications at any
future experiment that produces charm and includes a RICH.

We have established the tagging of $s$ jets using leading,
high-momentum identified $K^-$ and $\Lambda^0$ and produced the world's best
measurement of $A_s$.
Our studies of long-range correlations indicate that tagging of $u$ and $d$ jets
using leading particles is possible, and we hope to make a complete set of
measurements of the electroweak couplings of the $Z^0$ to the quarks.
The potential applications for light-flavor tagging are numerous, including
decays of $W^\pm$ bosons, top quarks, Higgs bosons, and new particles,
as well as in the study of jet production in deep inelastic scattering and
hadron-hadron collisions.

%\vfill
%\newpage

\section*{$^{**}$List of Authors} 
%
% author list for inclusion in LaTeX documents
% using \author{} and \address{} commands
%
% Institution number definitions:
%
\begin{center}
\def\iADEL{$^{(1)}$}
\def\iAOMORI{$^{(2)}$}
\def\iBOLO{$^{(3)}$}
\def\iBRUN{$^{(4)}$}
\def\iBU{$^{(5)}$}
\def\iCINC{$^{(6)}$}
\def\iCOLO{$^{(7)}$}
\def\iCOLU{$^{(8)}$}
\def\iCSU{$^{(9)}$}
\def\iFERR{$^{(10)}$}
\def\iFRAS{$^{(11)}$}
\def\iILLI{$^{(12)}$}
\def\iLBL{$^{(13)}$}
\def\iLTU{$^{(14)}$}
\def\iMASS{$^{(15)}$}
\def\iMISSI{$^{(16)}$}
\def\iMIT{$^{(17)}$}
\def\iMOSCOW{$^{(18)}$}
\def\iNAGO{$^{(19)}$}
\def\iOREG{$^{(20)}$}
\def\iOXF{$^{(21)}$}
\def\iPADO{$^{(22)}$}
\def\iPERU{$^{(23)}$}
\def\iPISA{$^{(24)}$}
\def\iRAL{$^{(25)}$}
\def\iRUTG{$^{(26)}$}
\def\iSLAC{$^{(27)}$}
\def\iSOGA{$^{(28)}$}
\def\iSOONG{$^{(29)}$}
\def\iTENN{$^{(30)}$}
\def\iTOHO{$^{(31)}$}
\def\iUCSB{$^{(32)}$}
\def\iUCSC{$^{(33)}$}
\def\iVAND{$^{(34)}$}
\def\iWASH{$^{(35)}$}
\def\iWISC{$^{(36)}$}
\def\iYALE{$^{(37)}$}

  \baselineskip=.75\baselineskip  
\mbox{K. Abe\unskip,\iAOMORI}
\mbox{K.  Abe\unskip,\iNAGO}
\mbox{T. Abe\unskip,\iSLAC}
\mbox{I.Adam\unskip,\iSLAC}
\mbox{T.  Akagi\unskip,\iSLAC}
\mbox{N. J. Allen\unskip,\iBRUN}
\mbox{A. Arodzero\unskip,\iOREG}
\mbox{W.W. Ash\unskip,\iSLAC}
\mbox{D. Aston\unskip,\iSLAC}
\mbox{K.G. Baird\unskip,\iMASS}
\mbox{C. Baltay\unskip,\iYALE}
\mbox{H.R. Band\unskip,\iWISC}
\mbox{M.B. Barakat\unskip,\iLTU}
\mbox{O. Bardon\unskip,\iMIT}
\mbox{T.L. Barklow\unskip,\iSLAC}
\mbox{J.M. Bauer\unskip,\iMISSI}
\mbox{G. Bellodi\unskip,\iOXF}
\mbox{R. Ben-David\unskip,\iYALE}
\mbox{A.C. Benvenuti\unskip,\iBOLO}
\mbox{G.M. Bilei\unskip,\iPERU}
\mbox{D. Bisello\unskip,\iPADO}
\mbox{G. Blaylock\unskip,\iMASS}
\mbox{J.R. Bogart\unskip,\iSLAC}
\mbox{B. Bolen\unskip,\iMISSI}
\mbox{G.R. Bower\unskip,\iSLAC}
\mbox{J. E. Brau\unskip,\iOREG}
\mbox{M. Breidenbach\unskip,\iSLAC}
\mbox{W.M. Bugg\unskip,\iTENN}
\mbox{D. Burke\unskip,\iSLAC}
\mbox{T.H. Burnett\unskip,\iWASH}
\mbox{P.N. Burrows\unskip,\iOXF}
\mbox{A. Calcaterra\unskip,\iFRAS}
\mbox{D.O. Caldwell\unskip,\iUCSB}
\mbox{D. Calloway\unskip,\iSLAC}
\mbox{B. Camanzi\unskip,\iFERR}
\mbox{M. Carpinelli\unskip,\iPISA}
\mbox{R. Cassell\unskip,\iSLAC}
\mbox{R. Castaldi\unskip,\iPISA}
\mbox{A. Castro\unskip,\iPADO}
\mbox{M. Cavalli-Sforza\unskip,\iUCSC}
\mbox{A. Chou\unskip,\iSLAC}
\mbox{E. Church\unskip,\iWASH}
\mbox{H.O. Cohn\unskip,\iTENN}
\mbox{J.A. Coller\unskip,\iBU}
\mbox{M.R. Convery\unskip,\iSLAC}
\mbox{V. Cook\unskip,\iWASH}
\mbox{R. Cotton\unskip,\iBRUN}
\mbox{R.F. Cowan\unskip,\iMIT}
\mbox{D.G. Coyne\unskip,\iUCSC}
\mbox{G. Crawford\unskip,\iSLAC}
\mbox{C.J.S. Damerell\unskip,\iRAL}
\mbox{M. N. Danielson\unskip,\iCOLO}
\mbox{M. Daoudi\unskip,\iSLAC}
\mbox{N. de Groot\unskip,\iSLAC}
\mbox{R. Dell'Orso\unskip,\iPERU}
\mbox{P.J. Dervan\unskip,\iBRUN}
\mbox{R. de Sangro\unskip,\iFRAS}
\mbox{M. Dima\unskip,\iCSU}
\mbox{A. D'Oliveira\unskip,\iCINC}
\mbox{D.N. Dong\unskip,\iMIT}
\mbox{P.Y.C. Du\unskip,\iTENN}
\mbox{R. Dubois\unskip,\iSLAC}
\mbox{B.I. Eisenstein\unskip,\iILLI}
\mbox{V. Eschenburg\unskip,\iMISSI}
\mbox{E. Etzion\unskip,\iWISC}
\mbox{S. Fahey\unskip,\iCOLO}
\mbox{D. Falciai\unskip,\iFRAS}
\mbox{C. Fan\unskip,\iCOLO}
\mbox{J.P. Fernandez\unskip,\iUCSC}
\mbox{M.J. Fero\unskip,\iMIT}
\mbox{K.Flood\unskip,\iMASS}
\mbox{R. Frey\unskip,\iOREG}
\mbox{T. Gillman\unskip,\iRAL}
\mbox{G. Gladding\unskip,\iILLI}
\mbox{S. Gonzalez\unskip,\iMIT}
\mbox{E.L. Hart\unskip,\iTENN}
\mbox{J.L. Harton\unskip,\iCSU}
\mbox{A. Hasan\unskip,\iBRUN}
\mbox{K. Hasuko\unskip,\iTOHO}
\mbox{S. J. Hedges\unskip,\iBU}
\mbox{S.S. Hertzbach\unskip,\iMASS}
\mbox{M.D. Hildreth\unskip,\iSLAC}
\mbox{J. Huber\unskip,\iOREG}
\mbox{M.E. Huffer\unskip,\iSLAC}
\mbox{E.W. Hughes\unskip,\iSLAC}
\mbox{X.Huynh\unskip,\iSLAC}
\mbox{H. Hwang\unskip,\iOREG}
\mbox{M. Iwasaki\unskip,\iOREG}
\mbox{D. J. Jackson\unskip,\iRAL}
\mbox{P. Jacques\unskip,\iRUTG}
\mbox{J.A. Jaros\unskip,\iSLAC}
\mbox{Z.Y. Jiang\unskip,\iSLAC}
\mbox{A.S. Johnson\unskip,\iSLAC}
\mbox{J.R. Johnson\unskip,\iWISC}
\mbox{R.A. Johnson\unskip,\iCINC}
\mbox{T. Junk\unskip,\iSLAC}
\mbox{R. Kajikawa\unskip,\iNAGO}
\mbox{M. Kalelkar\unskip,\iRUTG}
\mbox{Y. Kamyshkov\unskip,\iTENN}
\mbox{H.J. Kang\unskip,\iRUTG}
\mbox{I. Karliner\unskip,\iILLI}
\mbox{H. Kawahara\unskip,\iSLAC}
\mbox{Y. D. Kim\unskip,\iSOGA}
\mbox{R. King\unskip,\iSLAC}
\mbox{M.E. King\unskip,\iSLAC}
\mbox{R.R. Kofler\unskip,\iMASS}
\mbox{N.M. Krishna\unskip,\iCOLO}
\mbox{R.S. Kroeger\unskip,\iMISSI}
\mbox{M. Langston\unskip,\iOREG}
\mbox{A. Lath\unskip,\iMIT}
\mbox{D.W.G. Leith\unskip,\iSLAC}
\mbox{V. Lia\unskip,\iMIT}
\mbox{C.-J. S. Lin\unskip,\iSLAC}
\mbox{X. Liu\unskip,\iUCSC}
\mbox{M.X. Liu\unskip,\iYALE}
\mbox{M. Loreti\unskip,\iPADO}
\mbox{A. Lu\unskip,\iUCSB}
\mbox{H.L. Lynch\unskip,\iSLAC}
\mbox{J. Ma\unskip,\iWASH}
\mbox{G. Mancinelli\unskip,\iRUTG}
\mbox{S. Manly\unskip,\iYALE}
\mbox{G. Mantovani\unskip,\iPERU}
\mbox{T.W. Markiewicz\unskip,\iSLAC}
\mbox{T. Maruyama\unskip,\iSLAC}
\mbox{H. Masuda\unskip,\iSLAC}
\mbox{E. Mazzucato\unskip,\iFERR}
\mbox{A.K. McKemey\unskip,\iBRUN}
\mbox{B.T. Meadows\unskip,\iCINC}
\mbox{G. Menegatti\unskip,\iFERR}
\mbox{R. Messner\unskip,\iSLAC}
\mbox{P.M. Mockett\unskip,\iWASH}
\mbox{K.C. Moffeit\unskip,\iSLAC}
\mbox{T.B. Moore\unskip,\iYALE}
\mbox{M.Morii\unskip,\iSLAC}
\mbox{D. Muller\unskip,\iSLAC}
\mbox{V.Murzin\unskip,\iMOSCOW}
\mbox{T. Nagamine\unskip,\iTOHO}
\mbox{S. Narita\unskip,\iTOHO}
\mbox{U. Nauenberg\unskip,\iCOLO}
\mbox{H. Neal\unskip,\iSLAC}
\mbox{M. Nussbaum\unskip,\iCINC}
\mbox{N.Oishi\unskip,\iNAGO}
\mbox{D. Onoprienko\unskip,\iTENN}
\mbox{L.S. Osborne\unskip,\iMIT}
\mbox{R.S. Panvini\unskip,\iVAND}
\mbox{H. Park\unskip,\iOREG}
\mbox{C. H. Park\unskip,\iSOONG}
\mbox{T.J. Pavel\unskip,\iSLAC}
\mbox{I. Peruzzi\unskip,\iFRAS}
\mbox{M. Piccolo\unskip,\iFRAS}
\mbox{L. Piemontese\unskip,\iFERR}
\mbox{E. Pieroni\unskip,\iPISA}
\mbox{K.T. Pitts\unskip,\iOREG}
\mbox{R.J. Plano\unskip,\iRUTG}
\mbox{R. Prepost\unskip,\iWISC}
\mbox{C.Y. Prescott\unskip,\iSLAC}
\mbox{G.D. Punkar\unskip,\iSLAC}
\mbox{J. Quigley\unskip,\iMIT}
\mbox{B.N. Ratcliff\unskip,\iSLAC}
\mbox{T.W. Reeves\unskip,\iVAND}
\mbox{J. Reidy\unskip,\iMISSI}
\mbox{P.L. Reinertsen\unskip,\iUCSC}
\mbox{P.E. Rensing\unskip,\iSLAC}
\mbox{L.S. Rochester\unskip,\iSLAC}
\mbox{P.C. Rowson\unskip,\iCOLU}
\mbox{J.J. Russell\unskip,\iSLAC}
\mbox{O.H. Saxton\unskip,\iSLAC}
\mbox{T. Schalk\unskip,\iUCSC}
\mbox{R.H. Schindler\unskip,\iSLAC}
\mbox{B.A. Schumm\unskip,\iUCSC}
\mbox{J. Schwiening\unskip,\iSLAC}
\mbox{S. Sen\unskip,\iYALE}
\mbox{V.V. Serbo\unskip,\iWISC}
\mbox{M.H. Shaevitz\unskip,\iCOLU}
\mbox{J.T. Shank\unskip,\iBU}
\mbox{G. Shapiro\unskip,\iLBL}
\mbox{D.J. Sherden\unskip,\iSLAC}
\mbox{K. D. Shmakov\unskip,\iTENN}
\mbox{C. Simopoulos\unskip,\iSLAC}
\mbox{N.B. Sinev\unskip,\iOREG}
\mbox{S.R. Smith\unskip,\iSLAC}
\mbox{M. B. Smy\unskip,\iCSU}
\mbox{J.A. Snyder\unskip,\iYALE}
\mbox{H. Staengle\unskip,\iCSU}
\mbox{A. Stahl\unskip,\iSLAC}
\mbox{P. Stamer\unskip,\iRUTG}
\mbox{R. Steiner\unskip,\iADEL}
\mbox{H. Steiner\unskip,\iLBL}
\mbox{M.G. Strauss\unskip,\iMASS}
\mbox{D. Su\unskip,\iSLAC}
\mbox{F. Suekane\unskip,\iTOHO}
\mbox{A. Sugiyama\unskip,\iNAGO}
\mbox{S. Suzuki\unskip,\iNAGO}
\mbox{M. Swartz\unskip,\iSLAC}
\mbox{A. Szumilo\unskip,\iWASH}
\mbox{T. Takahashi\unskip,\iSLAC}
\mbox{F.E. Taylor\unskip,\iMIT}
\mbox{J. Thom\unskip,\iSLAC}
\mbox{E. Torrence\unskip,\iMIT}
\mbox{N. K. Toumbas\unskip,\iSLAC}
\mbox{A.I. Trandafir\unskip,\iMASS}
\mbox{J.D. Turk\unskip,\iYALE}
\mbox{T. Usher\unskip,\iSLAC}
\mbox{C. Vannini\unskip,\iPISA}
\mbox{J. Va'vra\unskip,\iSLAC}
\mbox{E. Vella\unskip,\iSLAC}
\mbox{J.P. Venuti\unskip,\iVAND}
\mbox{R. Verdier\unskip,\iMIT}
\mbox{P.G. Verdini\unskip,\iPISA}
\mbox{S.R. Wagner\unskip,\iSLAC}
\mbox{D. L. Wagner\unskip,\iCOLO}
\mbox{A.P. Waite\unskip,\iSLAC}
\mbox{Walston, S.\unskip,\iOREG}
\mbox{J.Wang\unskip,\iSLAC}
\mbox{C. Ward\unskip,\iBRUN}
\mbox{S.J. Watts\unskip,\iBRUN}
\mbox{A.W. Weidemann\unskip,\iTENN}
\mbox{E. R. Weiss\unskip,\iWASH}
\mbox{J.S. Whitaker\unskip,\iBU}
\mbox{S.L. White\unskip,\iTENN}
\mbox{F.J. Wickens\unskip,\iRAL}
\mbox{B. Williams\unskip,\iCOLO}
\mbox{D.C. Williams\unskip,\iMIT}
\mbox{S.H. Williams\unskip,\iSLAC}
\mbox{S. Willocq\unskip,\iSLAC}
\mbox{R.J. Wilson\unskip,\iCSU}
\mbox{W.J. Wisniewski\unskip,\iSLAC}
\mbox{J. L. Wittlin\unskip,\iMASS}
\mbox{M. Woods\unskip,\iSLAC}
\mbox{G.B. Word\unskip,\iVAND}
\mbox{T.R. Wright\unskip,\iWISC}
\mbox{J. Wyss\unskip,\iPADO}
\mbox{R.K. Yamamoto\unskip,\iMIT}
\mbox{J.M. Yamartino\unskip,\iMIT}
\mbox{X. Yang\unskip,\iOREG}
\mbox{J. Yashima\unskip,\iTOHO}
\mbox{S.J. Yellin\unskip,\iUCSB}
\mbox{C.C. Young\unskip,\iSLAC}
\mbox{H. Yuta\unskip,\iAOMORI}
\mbox{G. Zapalac\unskip,\iWISC}
\mbox{R.W. Zdarko\unskip,\iSLAC}
\mbox{J. Zhou\unskip.\iOREG}

\it
  \vskip \baselineskip                   % \bigskip did not work
%  \centerline{(The SLD Collaboration)}   % include collaboration name
%  \vskip \baselineskip        
  \baselineskip=.75\baselineskip   % shrink the interline spacing
\iADEL
  Adelphi University,
  South Avenue, Garden City, NY  11530 \break
\iAOMORI
  Aomori University,
  2-3-1 Kohata, Aomori City, 030 Japan \break
\iBOLO
  INFN Sezione di Bologna,
  Via Irnerio 46, I-40126 Bologna, Italy \break
\iBRUN
  Brunel University,
  Uxbridge, Middlesex, UB8 3PH United Kingdom \break
\iBU
  Boston University,
  590 Commonwealth Ave., Boston, MA  02215 \break
\iCINC
  University of Cincinnati,
  Cincinnati, OH  45221 \break
\iCOLO
  University of Colorado,
  Campus Box 390, Boulder, CO  80309 \break
\iCOLU
  Columbia University,
  Nevis Laboratories, P.O. Box 137, Irvington, NY  10533 \break
\iCSU
  Colorado State University,
  Ft. Collins, CO  80523 \break
\iFERR
  INFN Sezione di Ferrara,
  Via Paradiso 12, I-44100 Ferrara, Italy \break
\iFRAS
  Lab. Nazionali di Frascati,
  Casella Postale 13, I-00044 Frascati, Italy \break
\iILLI
  University of Illinois,
  1110 West Green St., Urbana, IL  61801 \break
\iLBL
  Lawrence Berkeley Laboratory,
  Dept. of Physics, 50B-5211 University of California, Berkeley, CA  94720 \break
\iLTU
  Louisiana Technical University,
  Ruston, LA  71272 \break
\iMASS
  University of Massachusetts,
  Amherst, MA  01003 \break
\iMISSI
  University of Mississippi,
  University, MS  38677 \break
\iMIT
  Massachusetts Institute of Technology,
  77 Massachussetts Avenue, Cambridge, MA  02139 \break
\iMOSCOW
  Moscow State University,
  Institute of Nuclear Physics, 119899 Moscow, Russia \break
\iNAGO
  Nagoya University,
  Nagoya 464, Japan \break
\iOREG
  University of Oregon,
  Department of Physics, Eugene, OR  97403 \break
\iOXF
  Oxford University,
  Oxford, OX1 3RH, United Kingdom \break
\iPADO
  Universita di Padova,
  Via F. Marzolo 8, I-35100 Padova, Italy \break
\iPERU
  Universita di Perugia, Sezione INFN,
  Via A. Pascoli, I-06100 Perugia, Italy \break
\iPISA
  INFN, Sezione di Pisa,
  Via Livornese 582/AS, Piero a Grado, I-56010 Pisa, Italy \break
\iRAL
  Rutherford Appleton Laboratory,
  Chilton, Didcot, Oxon, OX11 0QX United Kingdom \break
\iRUTG
  Rutgers University,
  Serin Physics Labs., Piscataway, NJ  08855 \break
\iSLAC
  Stanford Linear Accelerator Center,
  2575 Sand Hill Road, Menlo Park, CA  94025 \break
\iSOGA
  Sogang University,
  Ricci Hall, Seoul, Korea \break
\iSOONG
  Soongsil University,
  Seoul, Korea 156-743 \break
\iTENN
  University of Tennessee,
  401 A.H. Nielsen Physics Blg., Knoxville, TN  37996 \break
\iTOHO
  Tohoku University,
  Bubble Chamber Lab., Aramaki, Sendai 980, Japan \break
\iUCSB
  U.C. Santa Barbara,
  3019 Broida Hall, Santa Barbara,,CA ,93106 \break
\iUCSC
  U.C. Santa Cruz,
  Santa Cruz,,CA ,95064 \break
\iVAND
  Vanderbilt University,
  Stevenson Center, P.O.Box 1807, Station B, Nashville, TN  37235 \break
\iWASH
  University of Washington,
  Seattle, WA  98105 \break
\iWISC
  University of Wisconsin,
  1150 University Avenue, Madison, WI  53706 \break
\iYALE
  Yale University,
  5th Floor Gibbs Lab., P.O.Box 208121, New Haven, CT  06520  \break
\rm
%
%  }   % end of address list

\end{center}

\hfill
\end{document}